\documentclass[]{iopart}

\usepackage{graphicx,amssymb}

\begin{document}

\title{Eleven-orbit inspiral of a mass ratio 4:1 black-hole binary}

\author{U. Sperhake$^{1,2,3}$, B. Br{\"u}gmann$^{4}$, D. M{\"u}ller$^{4}$, 
C. F. Sopuerta$^1$}

\address{$^1$ Institute of Space Sciences (CSIC-IEEC), Campus UAB,
         Torre C5 parells, 08193 Bellaterra, SPAIN}

\address{$^2$ California Institute of Technology, 1200 E California Boulevard,
         Pasadena, CA 91125, USA}

\address{$^3$ Department of Physics and Astronomy,
         The University of Mississippi, University, MS 38677-1848, USA}

\address{$^4$~Theoretisch Physikalisches Institut, Friedrich Schiller
         Universit\"at, Max-Wien Platz 1, 07743 Jena, Germany}

\ead{sperhake@ieec.uab.es}

\begin{abstract}
  We analyse an eleven-orbit inspiral of a non-spinning black-hole
  binary with mass ratio $q\equiv M_1/M_2=4$. The numerically obtained
  gravitational waveforms are compared with post-Newtonian (PN)
  predictions including several sub-dominant multipoles up to
  multipolar indices $(l=5,m=5)$. We find that (i) numerical
  and post-Newtonian predictions of the phase of the $(2,2)$ mode
  accumulate a phase difference of about $0.35~{\rm rad}$ at
  the PN cut off frequency $M\omega=0.1$ for the Taylor T1
  approximant when numerical and PN waveforms are matched over
  a window in the early inspiral phase;
  (ii) in contrast to previous studies of equal-mass
  and specific spinning binaries, we find the Taylor T4
  approximant to agree less well with numerical results, {\em provided}
  the latter are extrapolated to infinite extraction radius;
  (iii) extrapolation
  of gravitational waveforms to infinite extraction radius is
  particularly important for subdominant multipoles
  with $l\ne m$; (iv) 3PN terms in post-Newtonian multipole
  expansions significantly improve the agreement with numerical
  predictions for sub-dominant multipoles.
\end{abstract}



The research area of Gravitational Wave (GW) Physics has acquired enormous
momentum in the course of the last decade. The ground-based detectors
LIGO, VIRGO and GEO600 have operated at design sensitivity
and the former two are currently being upgraded to advanced status
\cite{Abadie2010a, advancedVIRGO}.
Simultaneously, the planned ESA/NASA space mission LISA will soon enter a
crucial stage with the launch of the precursor mission LISA Pathfinder
\cite{Armano2009}.
Progress in theoretical GW source modelling has mirrored
that on the experimental side. Most importantly, the general
relativistic two-body problem for comparable-mass systems
has been solved using numerical relativity (NR)
techniques
\cite{Pretorius2005a, Campanelli2006, Baker2006},
leading to a wealth of insight into the dynamics of black-hole binaries
\cite{Pretorius2007a, Centrella2010}.

In order to maximise the scientific output of the expected
gravitational wave observations, it is crucial to have available
catalogues of highly accurate gravitational waveform templates
which are then used via the {\em matched filtering} technique
\cite{Finn1992}
to dig out physical signals from the observed data stream.
The construction of such waveform templates currently follows
either of the two following strategies. {\em Phenomenological
template banks} are based on {\em hybrid waveforms}
combining post-Newtonian predictions for most of the inspiral
with numerical relativity results for late inspiral, merger and ringdown.
Phase and amplitude are then assumed to be well-approximated by relatively
simple functions of the binary parameters and the GW frequency
$\omega$. Free parameters in these functions are determined by
comparison with a finite number of hybrid waveforms. This approach,
initially presented in \cite{Ajith2007, Ajith2007a, Ajith2007b}
has been applied to
a subset of spinning configurations in \cite{Ajith2009, Santamaria2010,
Sturani2010, Sturani2010a}.
The second approach, the {\em effective-one-body method} (EOB)
\cite{Buonanno1999}, targets
at semi-analytic predictions for the entire waveform via matching
post-Newtonian results to an effective-one-body metric and thus
models the binary through merger and ring-down.
Again, free parameters in the matching are determined by
comparison with a finite set of numerical relativity simulations.
Applications of the EOB method
can be found, for example, in
\cite{Damour2007a, Damour2007b, Damour2008, Boyle2008, Damour2010}
for non-spinning
and in \cite{Pan2009} for non-precessing spinning binaries.

Clearly, both approaches are heavily dependent on high-precision
numerical relativity results. Length requirements for the numerical
simulations in the context of gravitational wave detection
have been studied for non-spinning
binaries and special spin configurations in \cite{Hannam2010a} and
are in the range of about ten orbits. A recent study on
accuracy requirements for detection efficiency and parameter estimation
by MacDonald {\em et al.} \cite{MacDonald2011}
employs criteria developed in
\cite{Lindblom2008, Damour2010} and reports a larger number
of about 30 orbits required for hybridisation with current
post-Newtonian waveforms; for more details see these articles
and references therein.
Finally, we note that
numerical relativity waveforms are directly used in GW data analysis
inside the {\em Ninja} project \cite{Aylott2009, Aylott2009a}.

Purpose of our paper is to study in detail the accuracy of an eleven-orbit
inspiral of a non-spinning black-hole binary with mass ratio $q=4$
obtained with the {\sc Lean} code \cite{Sperhake2006}.
While higher-order multipoles have been investigated in the context
of the EOB model \cite{Buonanno2007a, Buonanno2009} and
PN-NR comparisons in \cite{Buonanno2006, Berti2007, Berti2007a, Campanelli2008},
we are not aware of their inclusion
in the construction of phenomenological models of
hybrid PN-NR waveforms. In this work, we will compare our
numerical results for the quadrupole as well as several subdominant modes
with the Taylor T1 and T4 approximants and different PN multipolar
expansions.

We start this paper with a brief summary of the numerical framework
in Sec.~\ref{sec:numerics}. The simulations and the numerical error
analysis is presented in Sec.~\ref{sec:runs}. In Sec.~\ref{sec:hybrid}
we discuss the different post-Newtonian approximants used in this
work and the method to construct hybrid waveforms. These are analysed
in Sec.~\ref{sec:analysis} and we conclude in Sec.~\ref{sec:conclusions}.


\section{Computational framework}
\label{sec:numerics}

The initial data for our black-hole binary configurations are
constructed according to the puncture method \cite{Brandt1997}.
Specifically, we use the analytic Bowen-York extrinsic
curvature \cite{Bowen1980} with linear momentum $\mathbf{P}$
and solve the Hamiltonian constraint for the conformal
factor using the spectral solver of Ansorg {\em et al.} \cite{Ansorg2004}.
In order to reduce the eccentricity of the binary system, we determine
the non-vanishing radial component $P_{\rm rad}$ of the initial
momentum via the iterative procedure described in Ref.~\cite{Pfeiffer2007}.

\begin{table}[t]
  \centering
  \caption{\label{tab:model}Initial bare mass parameters $m_1$, $m_2$, location
$x_1$, $x_2$ and tangential and radial linear momentum of black holes 1 and 2
of the binary. $v_{\rm kick}$ and $j_{\rm fin}$ are
the kick velocity and spin of the post-merger hole.}
  \begin{tabular}{cccccccc}
  \hline \hline
  $m_1/M$  &  $m_2/M$  &  $x_1/M$  &  $x_2/M$  &  $P_{\rm tan}/M$  &  $P_{\rm rad}/M$ & $v_{\rm kick}$ & $j_{\rm fin}$ \\
  \hline
  0.7923   &  0.1913  &  2.1865  &  $-8.746$  &  $\pm 0.05805$  &  $\mp 3.894\times 10^{-4}$ & $156.6~{\rm km/s}$ & $0.473$ \\
  \hline \hline
  \end{tabular}
\end{table}
The initial parameters thus obtained are given in units of the
total black-hole mass $M=M_1+M_2$ in Table \ref{tab:model}.
The bare mass parameters have been chosen such that the irreducible masses
as calculated with Thornburg's {\sc AHFinderDirect} \cite{Thornburg1996,
Thornburg2004} correspond exactly to a binary with mass ratio $q=4$.
For completeness we also give the final spin of the merged hole and
the recoil velocity. The latter is in excellent agreement with the
prediction $v_{\rm kick}=156.9~{\rm km/s}$ by \cite{Gonzalez2007},
confirming that the early inspiral does not significantly contribute to
the recoil; cf.~the discussion in Sec.~III C of \cite{Sperhake2010}.

These initial data are evolved in time with the so-called {\em moving
puncture} framework \cite{Campanelli2006, Baker2006} using the {\sc Lean}
code \cite{Sperhake2006} with upgrades to sixth-order spatial discretization
as summarised in Sec.~III of Ref.~\cite{Sperhake2007}. The {\sc Lean} code
is based on the {\sc Cactus} computational toolkit \cite{Cactusweb}
and employs mesh refinement provided by {\sc Carpet} \cite{Schnetter2003,
Carpetweb}. The exact implementation of the Baumgarte-Shapiro-Shibata-Nakamura
(BSSN) \cite{Shibata1995, Baumgarte1998}
formulation of the Einstein equations is given by Eqs.~(A1), (A4), (A6-A8)
of \cite{Sperhake2006}. We further impose a {\em floor} value
$\chi_{\rm floor}=10^{-4}$ on the conformal factor; cf.~\cite{Campanelli2006}.
The gauge variables are initialised as vanishing shift $\beta^i=0$
and a precollapsed lapse $\alpha=\sqrt{\chi}$, where
$\chi=(\det \gamma_{ij})^{-1/3}$ and $\gamma_{ij}$ is the three metric.
Lapse and shift are evolved according to Eqs.~(17) and (26)
of Ref.~\cite{vanMeter2006} with a damping parameter $\eta = 1.75/M$.

The computational domain consists of a set of five nested boxes centred
on the coordinate origin and five additional refinement levels with
two components each, centred on the black holes. In terms of the notation
in Sec.~II E of \cite{Sperhake2006}, the exact grid setup is given in units
of $M$ by
\begin{equation*}
  \{(307.2,\,153.6,\,102.4,\,32,\,16)\times (3.2,\,1.6,\,0.8,\,0.4,\,0.2), h_i\}
  \nonumber
\end{equation*}
with resolutions $h_0=M/160$, $h_1=M/180$, $h_2=M/200$, $h_3=M/220$ and
$h_4=M/240$ for studying the convergence properties. The lowest resolution
simulation using $h_0$ becomes unstable due to a gauge instability
at the outermost refinement boundary at about $t=1700~M$; see
\cite{Mueller2009, Schnetter2010,
Mueller2010, Alic2010} for more in-depth discussion and cures
of this instability. While we do not use this simulation directly
for comparison with post-Newtonian predictions, it extends
sufficiently far into the inspiral
to provide consistency checks of our convergence results.

Gravitational waves are extracted in the form of the Newman-Penrose scalar
$\Psi_4$ at radii $r_{\rm ex}=64~M$, $72~M$, $80~M$,
$88~M$, $96~M$, $104~M$ and $112~M$ using the method described in Appendix
C in \cite{Sperhake2006}. The resulting $\Psi_4$ is decomposed into multipoles
$\psi_{lm}$
by projection onto spherical harmonics of spin-weight $-2$ according to
Eq.~(2.1) of Ref.~\cite{Berti2007}; see also \cite{Bruegmann2006a} for
sign conventions.

Our analysis of the gravitational waves will use the gravitational
wave strain $h$ which is related to the Newman-Penrose scalar by
\begin{equation}
  \Psi_4 = \ddot{h}_+ - i\ddot{h}_{\times}.
\end{equation}
The multipoles $H_{lm}$ of the strain are related to those of $\Psi_4$
by Eq.~(II.5) of Brown {\em et al.} \cite{Brown2007a}, that is,
two integrations in time of $\psi_{lm}$. These integrations in time
represent a non-trivial operation and often result in low frequency
drifts \cite{Berti2007, Reisswig2010}. In order to circumvent these
problems, we use a modified version of the integration in Fourier space
as originally introduced in Ref.~\cite{Vaishnav2007}, see
also \cite{Campanelli2008}: the
Fourier transform $\bar{H}_{lm}(\omega)$ is divided by $-\omega^2$,
but set to zero inside a specified window. Our modification consists
in smoothing the filter function from Heaviside shape to
\begin{eqnarray}
  g(x) &=& \left\{
           \begin{array}{ll}
             0 & x < 0 \\
             \frac{1}{\mathcal{N}} \int x^4(1-x)^4dx \qquad & 0 < x < 1 \\
             1 & x > 1
           \end{array}
           \right.
\end{eqnarray}
where $x=(\omega - \omega_0)/\Delta \omega$ and $\mathcal{N}$ is
a normalisation constant. For the present study, we empirically find
$M\omega_0=0.001$ and $M\,\Delta \omega=0.001$ to provide
satisfactory results.
\begin{figure}[t]
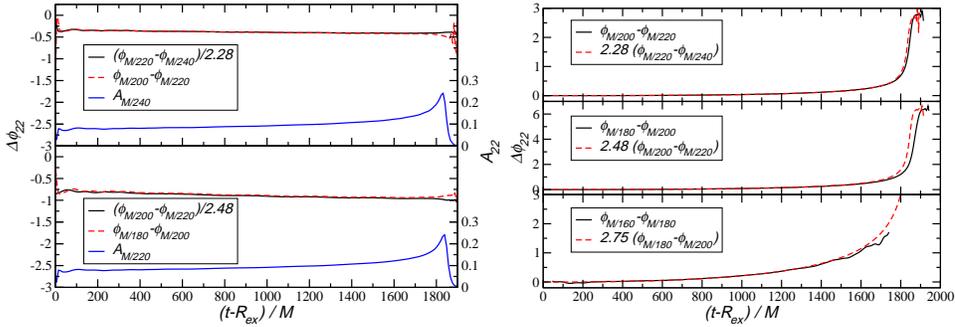

\centering
\includegraphics[height=4.3cm,clip=true]{f1a_conv_Amax.eps}
\includegraphics[height=4.3cm,clip=true]{f1b_conv_t0.eps}
\caption{Convergence of the phase $\phi_{22}$ for resolutions
         $h_1$, $h_2$, $h_3$ and $h_4$ and aligning the
         waveforms at the maximum of the $(2,2)$ mode (left panel)
         and for resolutions $h_0$, $h_1$, $h_2$, $h_3$ and $h_4$
         and aligning the waveforms at the start of the numerical
         simulation (right panel). The scaling factors correspond
         to eighth-order convergence. The lower solid curve in the
         left panel represents
         the amplitude $A_{22}$ and serves as orientation.}
\label{fig:conv}
\end{figure}
%

\section{Numerical results}
\label{sec:runs}

Before analysing the physical properties of the binary evolution,
we estimate the uncertainties due to discretization and wave
extraction at finite radii. The convergence of the phase $\phi$
obtained from the $l=2$, $m=2$ multipole is shown in Fig.~\ref{fig:conv}
for all resolutions $h_i$, $i=0,\ldots,4$. For the results
in the left panel, we have aligned
the waveforms in time at $t_{A_{22}}$, the time of
the peak of the amplitude of the (2,2) multipole.
Those displayed in the right panel are obtained for aligning the
waveforms at $t_0$, the start of the
simulation\footnote{The lowest resolution simulation
has only been used in the latter case because it does not extend
to the maximum of the amplitude of the (2,2) mode.}. The results are
consistent with eighth-order convergence for all resolutions
employed.
Similarly, we observe eighth-order convergence for
the amplitude. Bearing in mind that most of the discretization in our code is
sixth-order, it appears that the leading-order discretization errors
are subdominant. Whether this occurs due to systematic cancellation
is hard to identify because of the
enormous complexity of the Einstein equations. In order to test whether
the observed convergence is coincidental or systematic, we have performed
the larger set of runs employing five different resolutions and using
different choices of aligning the waveforms in time.
The observation of the same clean convergence order for all cases
demonstrates the robustness of our observations. For the determination
of uncertainties due to discretization, however, we will calculate
Richardson extrapolated results using a more conservative
sixth-order extrapolation.

\begin{figure}[t]
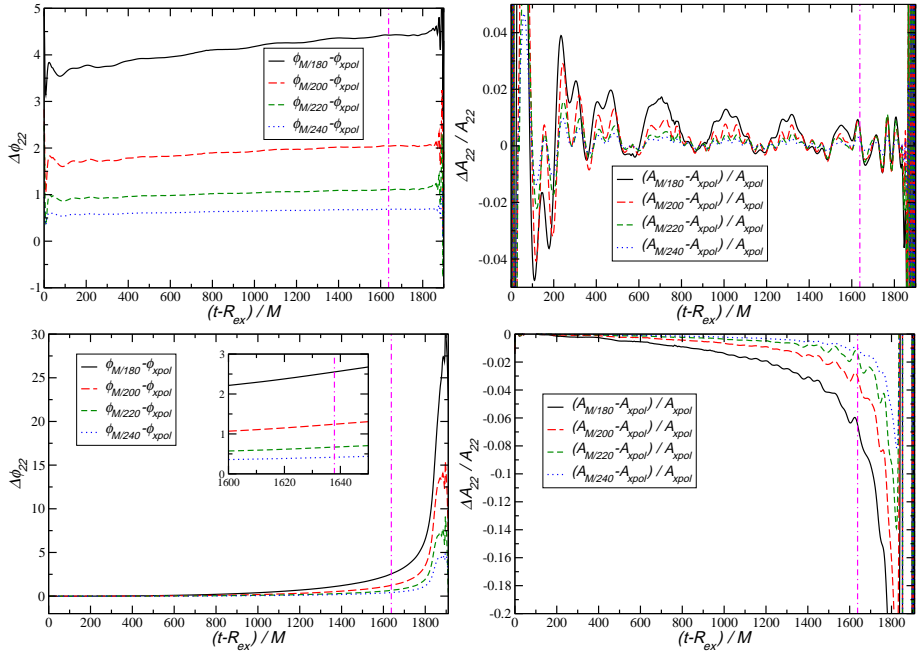

\centering
\includegraphics[height=4.3cm,clip=true]{f2a_seq_phi_h.eps}
\includegraphics[height=4.3cm,clip=true]{f2b_seq_A_h.eps}
\includegraphics[height=4.3cm,clip=true]{f2c_seq_phit0_h.eps}
\includegraphics[height=4.3cm,clip=true]{f2d_seq_At0_h.eps}
\caption{Deviation of the phase $\phi$ (left) and relative deviation
         of the amplitude (right panel) obtained for finite resolution
         from the Richardson extrapolated values and for
         aligning the waveforms at $t_{A_{22}}$, the maximum of the
         amplitude of the (2,2) multipole (upper panels) and at
         the start of the simulation $t_0$
         (lower panels). The vertical lines
         line marks the time $t_{\omega}$
         where the frequency of the (2,2) multipole
         reaches $M\omega_{22}=0.1$.}
\label{fig:dAphi_h}
\end{figure}

The deviations of phase and amplitude obtained at
finite resolutions from the Richardson extrapolated values
are shown in Fig.~\ref{fig:dAphi_h} for both types of alignment
of the waveforms, at $t_{A_{22}}$
(upper panels) and at $t_0$
(lower panels). For the former alignment,
the figure implies a phase error of
$\Delta \phi \lesssim 0.6~{\rm rad}$ for the high-resolution simulation.
We note, however, that the phase difference between the high resolution
and the extrapolated result is nearly constant until the
late ringdown stage. Because a constant phase offset does not
enter the comparison with PN results, the relevant phase error for this
purpose is given by the variation of the difference over the simulation
which is significantly smaller, about $\Delta \phi \approx 0.11~{\rm rad}$.
The relative amplitude error shown in Fig.~\ref{fig:dAphi_h} is
less than $1~\%$ and, as we shall see below, dominated by the uncertainty
arising from the use of finite extraction radii.

In case of aligning the waveforms at $t_0$,
Fig.~\ref{fig:dAphi_h} implies an accumulated phase error of about
0.4~rad for the phase of the high resolution run
at $t_{\omega}$ defined as the time when the frequency of the multipole
reaches $M\omega_{22}=0.1$, the endpoint of the post-Newtonian integration
in Sec.~\ref{sec:hybrid} (vertical lines in the figure).
The amplitude error has grown to $2~\%$ at the same time.
We note here that alignment of the gravitational waves at $t_{A_{22}}$
results in smaller uncertainty estimates.
For the analysis of Sec.~\ref{sec:analysis},
however, estimates obtained for both type of alignments are adequate.
In that analysis we will exclusively use the high-resolution
data set.

In order to estimate the error due to finite extraction radius, we
assume a polynomial dependence of the phase error on $1/R_{\rm ex}$
\cite{Boyle2007}. While the linear term is expected to dominate the
error, in practise the quadratic term may also be significant
\cite{Hannam2007}. We therefore fit the numerical data assuming either
of
\begin{eqnarray}
  f(r) &=& f_0 + \frac{f_1}{r}, \label{eq:Rex1} \\
  f(r) &=& f_0 + \frac{f_1}{r} + \frac{f_2}{r^2}, \label{eq:Rex2}
\end{eqnarray}
and denote the extrapolated results thus obtained ``xpol1'' and ``xpol2'',
respectively. Eventually, we use the extrapolated result ``xpol1'' in the
\begin{figure}[t]
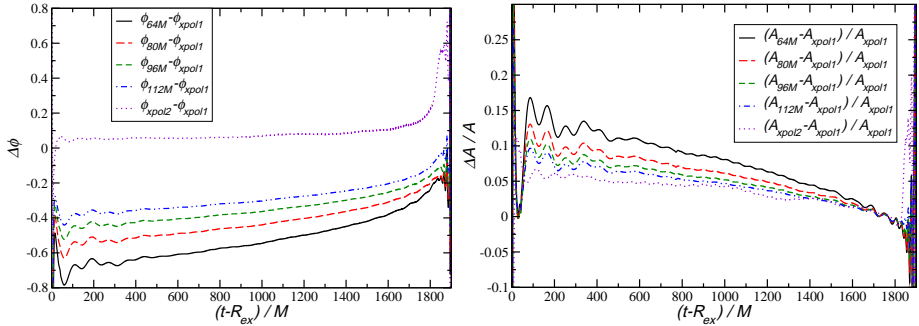

\centering
\includegraphics[height=4.3cm,clip=true]{f3a_dphi_Rex.eps}
\includegraphics[height=4.3cm,clip=true]{f3b_dA_Rex.eps}
\caption{Deviation of the phase $\phi$ (left) and relative deviation
         of the amplitude (right panel) obtained at finite extraction radius
         from the values extrapolated according to Eq.~(\ref{eq:Rex1}).}
\label{fig:dAphi_Rex}
\end{figure}
analysis below and estimate the error by its deviation from
the ``xpol2'' result. These deviations are shown
for phase and amplitude and a subset of all extraction radii in
Fig.~\ref{fig:dAphi_Rex}. From the figure we infer a relative error
of $5~\%$ or less for the amplitude while the phase error is
$\le 0.2~{\rm rad}$ up to merger and increases to about $0.5~{\rm rad}$
in the late ringdown stage.

The phase uncertainties of higher order multipoles behave similarly
to those of the quadrupole, but we find the overall errors to
scale approximately with the multipole index $m$. The amplitude
errors of subdominant modes are significantly larger, however,
due to the lower signal combined with numerical noise. Extrapolation
of the amplitude to infinite extraction radius amplifies the numerical
noise and we therefore present higher order multipoles using extrapolated
phase, but amplitude from the largest extraction radius instead. The
numerical uncertainties of these amplitudes are about $12~\%$ for the
$(3,3)$ mode, $20~\%$ for the $(4,4)$ and $(5,5)$ mode and $25~\%$ for the
$(3,2)$ and $(4,3)$ mode. While these errors are too large for a high precision
study, they will allow us to calibrate the significance of 3PN order
terms in the post-Newtonian multipole expansion in Sec.~\ref{sec:analysis}.

The errors stated so far are internal checks of accuracy. For the
purpose of an external verification of our results, we compare the
amplitude and phase of the $l=2,~m=2$ mode to results obtained with
the independent {\sc BAM} code~\cite{Bruegmann2004,Bruegmann2006a}. In
order to avoid additional uncertainties arising from the integration of
the Newman-Penrose scalar $\Psi_4$ to strain $h$, here we compare the
modes of $\Psi_4$ which we directly obtain from the numerical simulations
in both codes. The {\sc BAM} simulations also use lower resolutions
(hence the larger uncertainties) and different extraction radii. We
therefore present this comparison using results extrapolated both in
resolution, assuming second order convergence for the {\sc BAM} runs,
and extraction radius, using Eqs.~(\ref{eq:Rex1}), (\ref{eq:Rex2}).

The uncertainties of the {\sc BAM} simulation are studied in detail in
\cite{Hannam2010} and are summarised as follows. The accumulated
phase error for the $N=96$ high resolution
run at $t_{\omega}$ as obtained from Richardson
extrapolation and for aligning the waveforms at $t_0$ is 0.46~rad
and the error due to the use of finite
extraction radius at $R_{\rm ex}=90~M$ is 0.15~rad.
The corresponding uncertainty in the
amplitude is about $10~\%$ for both, discretization and finite
extraction radius during the inspiral and merger.

\begin{figure}[t]
\centering
\includegraphics[height=4.0cm,clip=true]{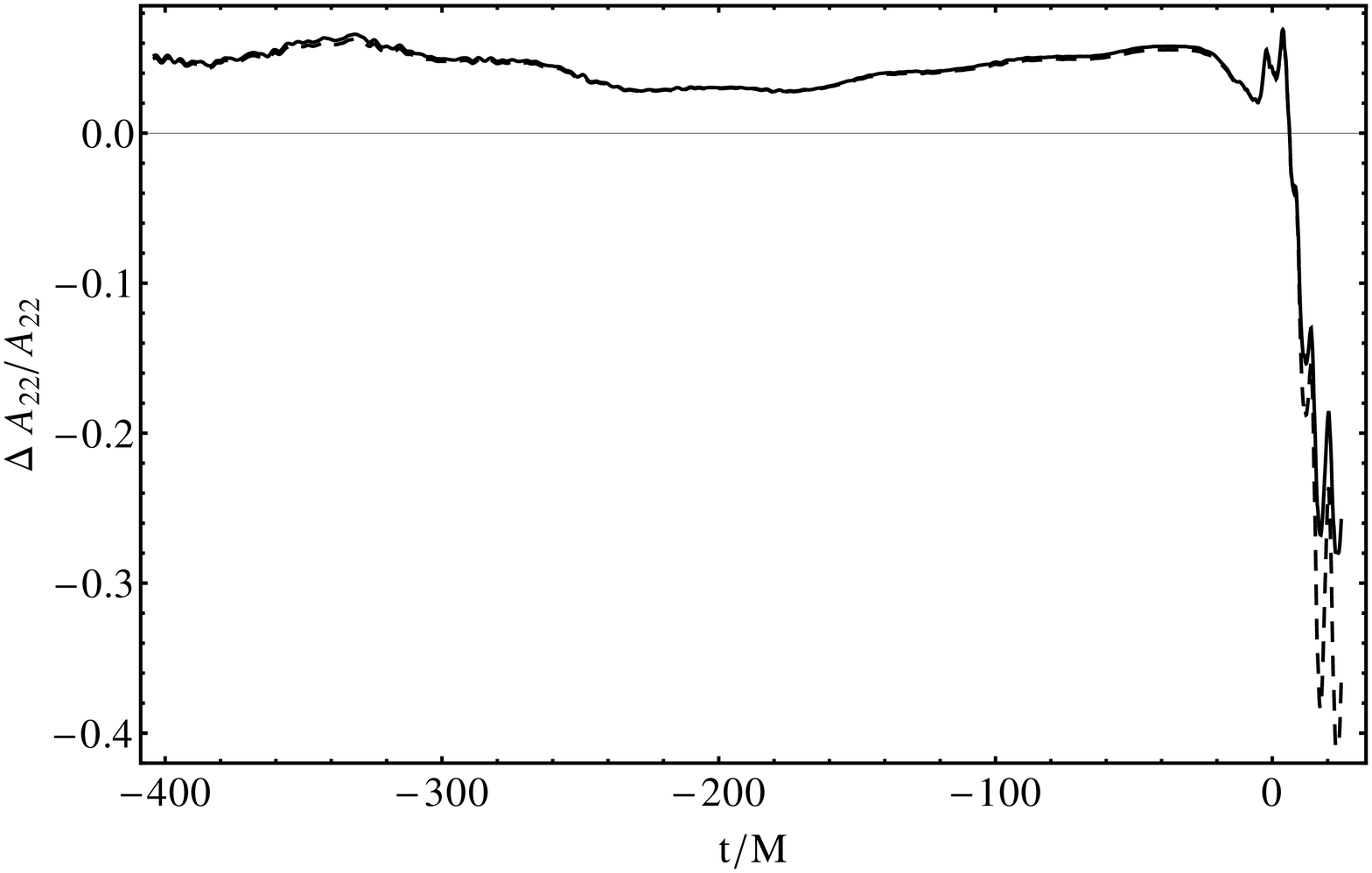}
\includegraphics[height=4.0cm,clip=true]{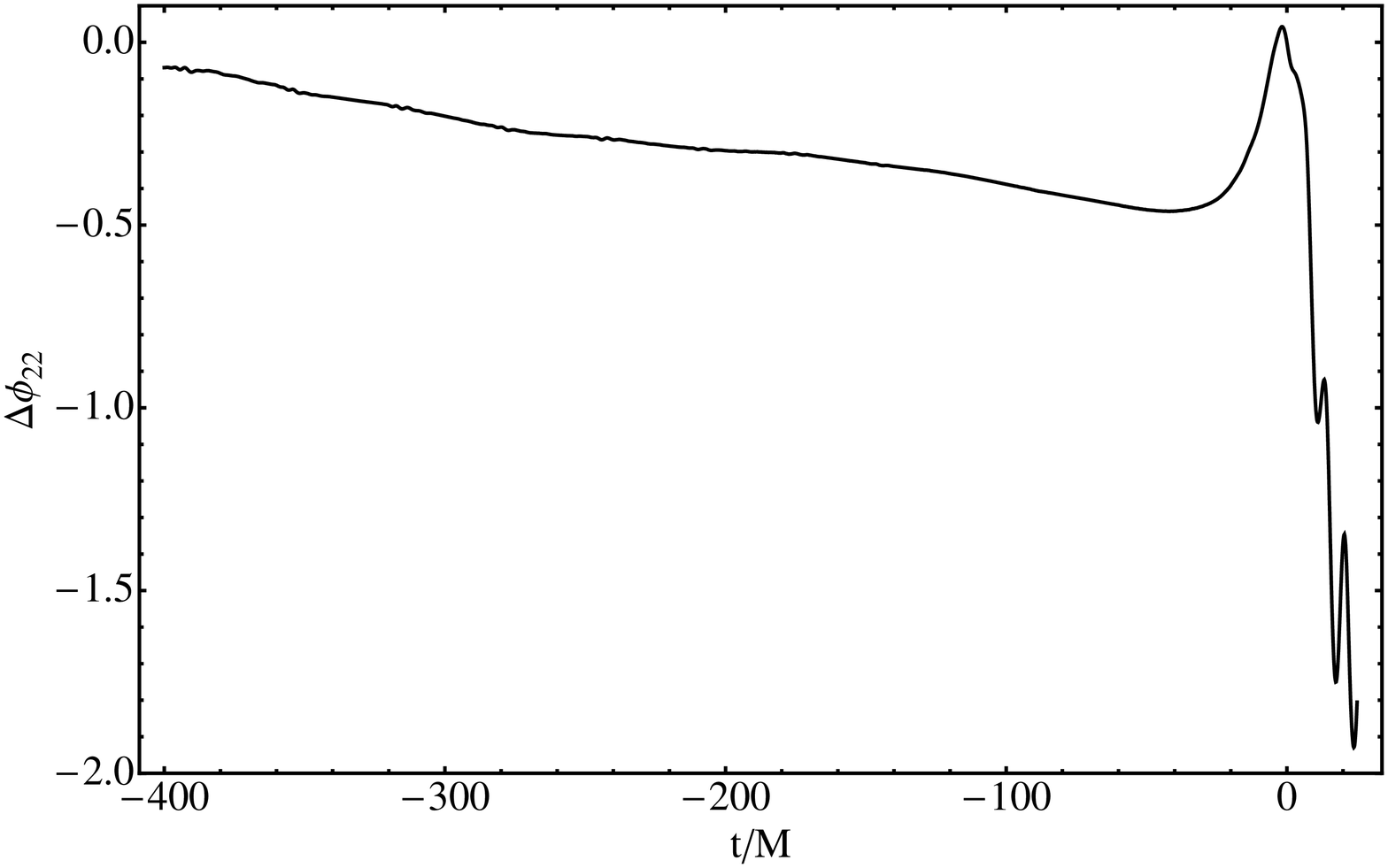}
\put(-300,20){\includegraphics[height=0.8cm]{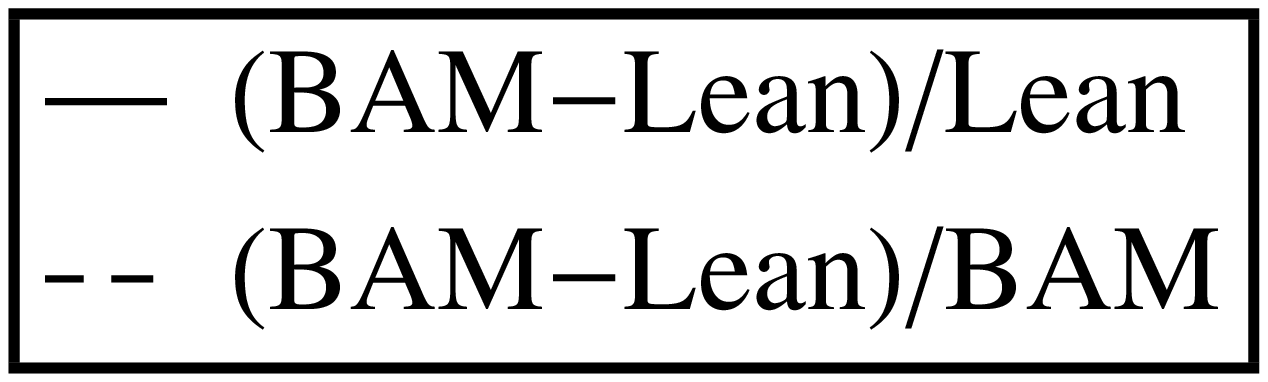}}
\caption{
  Relative amplitude differences (left panel) and phase differences (right
  panel) of the $l=2,~m=2$ modes of $\Psi_4$ between the extrapolated
  (in resolution and extraction radius) results from the {\sc BAM} and the
  {\sc Lean} codes.  Results from both codes are aligned at the time of peak
  amplitude $t_{A_{22}}$. We note that the relatively large percentage
  discrepancy between the results of the two codes at late times
  is a consequence of the very low amplitude of the wave signal
  in the late ringdown.
  }
\label{fig:LeanBamDiff}
\end{figure}

Relative deviations between the amplitudes and phase differences obtained
from the two codes are shown in Fig.~\ref{fig:LeanBamDiff}. Note that
the waveforms have been aligned at $t_{A_{22}}$ defined as
$t\equiv 0$ on the horizontal axes of this figure.
This alignment becomes necessary because of the different initial separations
(and, hence, duration) of the simulations, $D=10~M$ for {\sc BAM} and
$D=11~M$ for {\sc Lean}. Phase and amplitude differences shown in the figure
are well within the uncertainty estimates of the two simulations.

Finally, we evaluate the residual eccentricity of the binary simulated
with {\sc Lean} as a function
of time according to \cite{Mroue2010}
\begin{equation}
  e_{\phi}(t) = \frac{\phi_{\rm NR}(t) - \phi_{\rm fit}(t)}{4},
  \label{eq:ecc}
\end{equation}
where $\phi_{\rm NR}$ is the phase of the numerical $(2,2)$ mode and
$\phi_{\rm fit}$ a seventh order polynomial fit of this phase over
a time window $t-R_{\rm ex} = 250\ldots 1600~M$.
We thus obtain an oscillatory pattern of $e_{\phi}(t)$ with an amplitude
gradually decreasing from $0.005$ early on
towards $0.002$ one to two orbits before
merger.

\section{Post-Newtonian approximants and hybrid waveforms}
\label{sec:hybrid}

Post-Newtonian waveforms are calculated using the so-called {\em Taylor T1}
and {\em Taylor T4} approximants; cf.~\cite{Boyle2007} for a summary
including further approximants. Frequency and phase are obtained from
the system of ordinary differential equations
\begin{eqnarray}
  \frac{dx}{dt} &=& - \frac{\mathcal{L}}{dE/dx}, \label{eq:dxdt} \\
  \frac{d\Phi}{dt} &=& \frac{1}{M} x^{3/2}. \label{eq:dPhidt}
\end{eqnarray}
Here $x$ is related to the orbital frequency via $x\equiv (M\Omega)^{2/3}$,
$\Phi$ is the orbital phase, and the gravitational wave flux $\mathcal{L}$
and the orbital energy $E$ are given by Eqs.~(231) and (203), respectively,
of Blanchet's review \cite{Blanchet2006}. The difference between the
Taylor T1 and Taylor T4 approximant is the treatment of the right hand side
of Eq.~(\ref{eq:dxdt}). The former calculates the quotient
$\mathcal{L}/(dE/dx)$ numerically from the PN truncated
individual expressions for numerator and denominator while the T4 approximant
expands the quotient in a series in $x$ and truncates this series at the
appropriate PN order, 3.5 in our case.
The two approximants thus differ only at higher PN order, but previous studies
have found the Taylor T4 approximant to produce particularly good agreement
with numerical results \cite{Boyle2007, Hannam2010}.

In either case, we obtain the rescaled frequency $x$ and the orbital phase
$\Phi$ from integration of Eqs.~(\ref{eq:dxdt}) and (\ref{eq:dPhidt}).
The gravitational wave multipoles are given in terms of these
quantities by Kidder's
Eqs.~(79)-(116) in \cite{Kidder2007} (to 3PN order for $(2,2)$ and $(4,4)$
and 2.5PN order otherwise) or Eq.~(9.4) of the more recent study
by Blanchet {\em et al.} \cite{Blanchet2008} (to 3PN order).
Unless specified otherwise, results displayed use the latter 3PN terms.

In order to construct a hybrid waveform, we need to determine the integration
constants $\Phi_0$ and $t_0$ of the system
(\ref{eq:dxdt}), (\ref{eq:dPhidt}). In practise, this is achieved by maximising
the overlap between the real part of the post-Newtonian multipole
$H_{22,{\rm PN}}$ and its numerical counterpart $H_{22,{\rm NR}}$
over a matching window
$t_1\le t-R_{\rm ex} \le t_2~M$.
In this context, the overlap of two functions $f$ and $g$ is defined as
\begin{eqnarray}
  \xi &\equiv& \frac{ \langle f,g \rangle }{\sqrt{ \langle f,f \rangle
       \langle g,g \rangle }}, \\
  \langle f,g \rangle &\equiv& \int_{t_1}^{t_2} f(t)\,g(t)dt.
\end{eqnarray}
The maximisation is achieved using the downhill simplex method of
Nelder \& Mead \cite{Nelder1965,Press1992}.
Finally, we combine the PN and numerical waveform into a hybrid
according to
\begin{equation}
  H_{lm} = (1-w)\alpha H_{lm,{\rm PN}} + w H_{lm,{\rm NR}} \label{eq:hybrid},
\end{equation}
where the weighting function $w=1$ for $t<t_1$, $w=0$ for $t>t_2$ and
for values $t_1<t<t_2$ a smooth transition is given by
\begin{eqnarray}
  w &=& 630 \left( \frac{1}{9}z^9 - \frac{1}{2}z^8 + \frac{6}{7}z^7
        -\frac{2}{3}z^6 + \frac{1}{5} z^5 \right), \\
  z &\equiv& \frac{t-t_1}{t_2-t_1}.
\end{eqnarray}
The additional factor $\alpha$ has been introduced to compensate for
differences in the amplitude between the post-Newtonian and numerical
results. For each multipole it is chosen such that the average PN wave
amplitude inside the interval $[t_1,t_2]$ matches the NR result.

In order to test the robustness of our results versus the details
of the matching procedure, we perform an alternative matching
by equating phase and time at a fiducial point in time chosen to
be $t_{\omega}$ where the gravitational wave frequency of the (2,2) mode
takes the value $M\omega_{22}=0.1$.

\section{Analysis}
\label{sec:analysis}

\begin{figure}[t]
\centering
\includegraphics[height=5.50cm,clip=true]{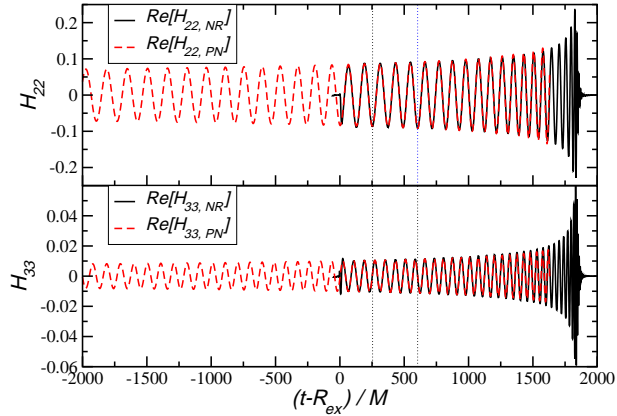}
\caption{Real part of the $l=2,~m=2$ and the $l=3,~m=3$ gravitational
         wave modes. The Taylor T1 PN waveforms (dashed) are matched to the
         numerical results in the window $250~M\le t-R_{\rm ex} \le 600~M$
         indicated by vertical dotted lines.}
\label{fig:PNNR}
\end{figure}

The results of the matching procedure using a window $t_1=250~M$,
$t_2=600~M$ are illustrated for the Taylor T1
approximant in
Fig.~\ref{fig:PNNR}, which shows the real part of the $(2,2)$ and the
$(3,3)$ modes. The matching window is indicated by the vertical
dotted lines. The corresponding figure for Taylor T4 would look nearly
identical.

\begin{figure}[t]
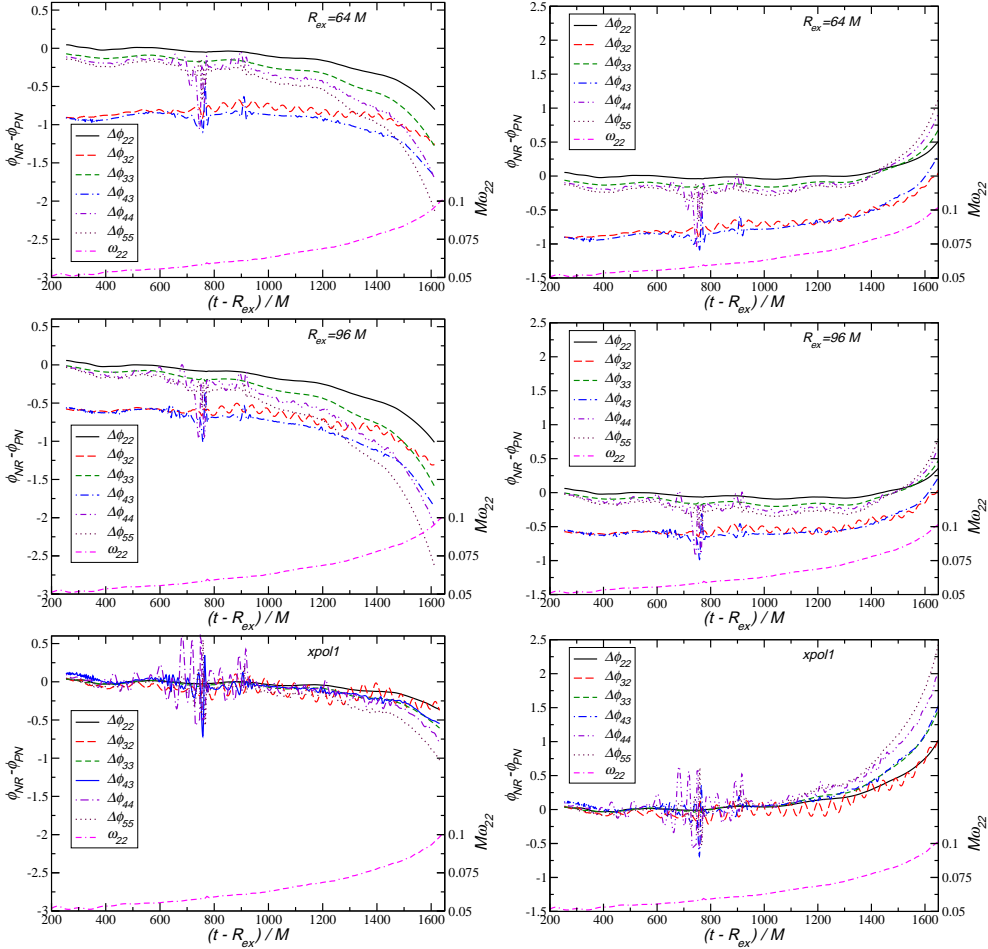

\centering
\includegraphics[height=4.18cm,clip=true]{f6a_T1_dphi_rex02.eps}
\includegraphics[height=4.18cm,clip=true]{f6b_T4_dphi_rex02.eps}
\includegraphics[height=4.18cm,clip=true]{f6c_T1_dphi_rex06.eps}
\includegraphics[height=4.18cm,clip=true]{f6d_T4_dphi_rex06.eps}
\includegraphics[height=4.18cm,clip=true]{f6e_T1_dphi_rexinf.eps}
\includegraphics[height=4.18cm,clip=true]{f6f_T4_dphi_rexinf.eps}
\caption{Phase difference between the PN and NR results obtained by using
         a matching window with $t_1=250~M$ and $t_2=600~M$ and
         numerical results at (from top to bottom) $R_{\rm ex}=64~M$,
         $96~M$ and extrapolated to infinity. Results in the left
         column are obtained for the Taylor T1 expansion, those on the
         right for Taylor T4. The dash-dash-dotted curve near the bottom
         of each panel gives the GW frequency of the $(2,2)$ mode for
         reference.}
\label{fig:dphiT1T4}
\end{figure}

In order to quantitatively analyse the agreement between the PN and NR
results, we display in Fig.~\ref{fig:dphiT1T4} the phase differences
for various multipoles obtained by using the Taylor T1 (left panels)
and T4 (right panels) approximants and numerical results extracted
at $R_{\rm ex}=64~M$ and $96~M$ as well as extrapolated to infinite
extraction radius. For orientation we plot near the bottom of each panel the
gravitational wave frequency $M\omega_{22}$ of the $l=2,~m=2$ mode
(dash-dash-dotted curve). All curves end at $M\omega_{22}=0.1$ which is
the maximum frequency chosen for the PN integration.

We begin our discussion with the dominant quadrupole mode $l=2$, $m=2$.
In all cases, the phase agreement between PN and NR results is better
than $\Delta \phi =0.1~{\rm rad}$ inside the matching window, but
gradually increases as the inspiral proceeds until the accumulated
phase discrepancy reaches values between $0.25~{\rm rad}$ and
$0.75~{\rm rad}$. At both finite extraction radii the agreement
grows to larger values for the Taylor T1 expansion than the T4 approximant.
The behaviour of the $(3,3)$, $(4,4)$ and $(5,5)$ modes is similar, although
with larger phase discrepancy approximately proportional to the
multipole index $m$. This is not too surprising bearing in mind that
the phase grows faster for higher-order modes with the same proportionality.

Those subdominant multipoles with $l\ne m$, however, exhibit a drastically
different behaviour when extracted at finite radius. For both, the
$(3,2)$ and the $(4,3)$ mode the numerical phase differs significantly from the
PN predictions throughout the entire simulation. We note, in this context,
that the orbital phase in the PN expressions is fixed exclusively using
\begin{figure}[t]
\centering
\includegraphics[height=5.50cm,clip=true]{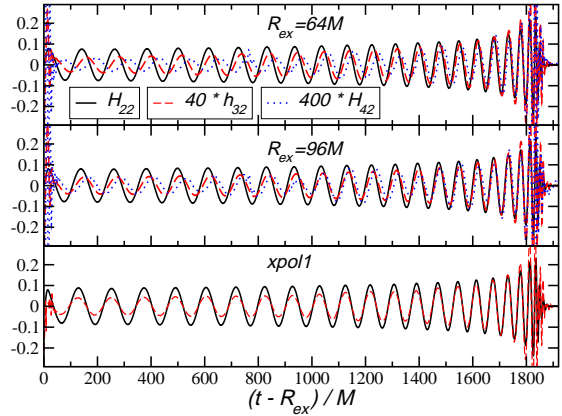}
\caption{The $(2,2)$ (solid), $(3,2)$ (dashed) and $(4,2)$ (dotted) multipole
         extracted at $R_{\rm ex}=64~M$ and $96~M$ (upper panels)
         and the former two
         extrapolated to infinite extraction radius (bottom panel).}
\label{fig:H223242}
\end{figure}
information from the $(2,2)$ mode. Furthermore, inspection of the post-Newtonian
multipoles in Refs.~\cite{Kidder2007} and \cite{Blanchet2008}
reveals that the $(2,2)$ and the $(3,2)$ GW multipoles are almost in
phase\footnote{The minor dephasing due to the complex PN amplitudes
of the $(2,2)$ and $(3,2)$ modes is negligible in this context.}. Indeed, this
is also the case for the numerical modes {\em provided} we extrapolate
to infinite extraction radius, as becomes apparent in the bottom panels
where the large phase error of the $(3,2)$ and $(4,3)$ modes disappear. We also
illustrate this feature in Fig.~\ref{fig:H223242}
where we plot the real part of the numerical $(2,2)$, $(3,2)$ and $(4,2)$
modes\footnote{The low amplitude of the $(4,2)$ combined with numerical noise
prevent us from performing accurate extrapolation to infinite
extraction radius which is also the reason it is not included in the
remainder of this study.}. The figure demonstrates that
(i) the dephasing is largest
for the $(4,2)$ mode, (ii) the dephasing decreases at larger extraction
radius and (iii) the dephasing virtually disappears if we extrapolate
to infinite extraction radius.
We make exactly the same observations for the $(4,3)$ and the $(3,3)$
\begin{figure}[t]
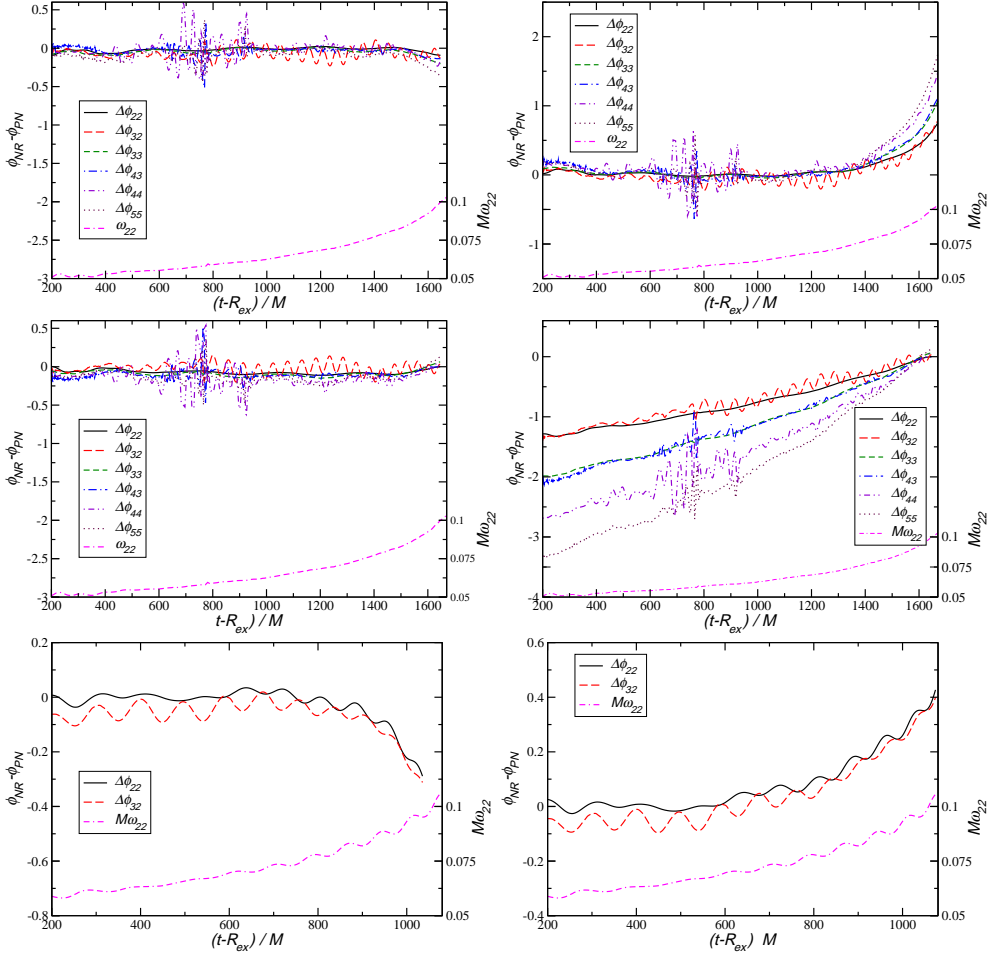

\centering
\includegraphics[height=4.20cm,clip=true]{f8a_dphiPNNR_T1.eps}
\includegraphics[height=4.20cm,clip=true]{f8b_dphiPNNR_T4.eps}
\includegraphics[height=4.20cm,clip=true]{f8c_dphiPNNR_omT1.eps}
\includegraphics[height=4.20cm,clip=true]{f8d_dphiPNNR_omT4.eps}
\includegraphics[height=4.20cm,clip=true]{f8e_dphiPNNR_bamT1.eps}
\includegraphics[height=4.20cm,clip=true]{f8f_dphiPNNR_bamT4.eps}
\caption{Phase difference between the PN and NR results obtained by using
         numerical results extrapolated to infinity and employing
         a later matching window $t1=900~M$, $t_2=1250~M$ (upper panels),
         matching phase and time at a fiducial point in time
         where $M\omega_{22}=0.1$ (centre panels) and using a matching
         window $t_1=250~M$, $t_2=600~M$ for (2,2) and (3,2)
         multipoles of the BAM simulation (bottom panels). Results in the left
         column are obtained for the Taylor T1 expansion, those on the
         right for Taylor T4. The dash-dash-dotted curve near the bottom
         of each panel gives the GW frequency of the $(2,2)$ mode for
         reference.}
\label{fig:dphiT1T4extra}
\end{figure}
multipoles;
see the short-dashed and dash-dotted curves in Fig.~\ref{fig:dphiT1T4}.
Our findings demonstrate the importance of having available either
reliable wave extraction tools at future null infinity
\cite{Reisswig2009, Reisswig2009a, Babiuc2010} or accurate extrapolations
from results at finite radii \cite{Boyle2009a}.

A further interesting feature of the extrapolated modes is the significant
improvement of the phase agreement between NR and the Taylor T1 approximant;
for $R_{\rm ex} \rightarrow \infty$,
Taylor T1 provides better agreement than T4.
We thus cannot confirm for our unequal-mass binary the exceptional
behaviour of the T4 approximant reported for the equal-mass case
by \cite{Boyle2007} and a $q=4$
unequal-mass binary in Fig.~8
of Ref.~\cite{Hannam2010}. We note, however, that this does not
constitute an incompatibility of our results with those of
Ref.~\cite{Hannam2010}, because the latter are based on waveforms
at finite extraction radius $90~M$, where we also obtain better agreement
of numerical results with T4.

We next perform the following tests in order to investigate
the robustness of our observations at infinite
extraction radius. (i) Motivated by the observation
of noise in the numerical phase due to spurious initial radiation
in the earlier part of their waveform
in Ref.~\cite{Boyle2008}, cf.~their Fig.~17\footnote{We note that
Boyle {\em et al.} use different initial data and that the
amplitude of the noise is at least one order of magnitude below
the effects investigated in our work, compatible with the absence
of such noise in the (2,2) mode on the scale of our Fig.~\ref{fig:dphiT1T4}.},
we choose a later matching window $t_1=900~M$, $t_2=1250~M$.
(ii) Instead of using a window, we match phase and time at the fiducial
point in time $t_{\omega}$ where $M\omega_{22}=0.1$ as done for example in
Ref.~\cite{Hannam2010}. (iii) We apply the matching procedure
with $t_1=250~M$, $t_2=600~M$ to the BAM waveform. The results
are shown, from top to bottom, in Fig.~\ref{fig:dphiT1T4extra}.
As expected, different alignment procedures produce different
functions of time for the phase differences. But for all cases,
we observe better agreement of the numerical simulations with
the Taylor T1 prediction as compared with Taylor T4. This is
also the case for the BAM simulation which is furthermore
consistent within the respective uncertainties with that
obtained with the {\sc Lean} code. We conclude that the exceptional
agreement of Taylor T4 observed for some specific black-hole
configurations is most likely explained by a coincidental cancellation
of higher-order post-Newtonian corrections which does not hold for general
systems.

Finally, we consider differences in the amplitude predictions
of post-Newtonian and numerical relativity results.
The PN expressions for the gravitational wave multipoles in
Refs.~\cite{Kidder2007} and \cite{Blanchet2008} differ in the inclusion
of higher order terms in several subdominant modes in the latter
work. In order to assess the significance of these higher-order
terms, we have performed the matching of numerical to PN waveforms
as described in Sec.~\ref{sec:hybrid} with $t_1=250~M$ and $t_2=600~M$
using either set of
multipole expressions. As mentioned above, the matching process
involves a rescaling of the PN waveforms to compensate for amplitude
differences in the matching window,~the factor $\alpha$ in
Eq.~(\ref{eq:hybrid}).
For all subdominant multipoles which have been extended to 3PN order in
\cite{Blanchet2008}, we observe in Table \ref{tab:alpha}
an improvement in the agreement between NR
and PN amplitudes, i.~e.~a value $\alpha$ closer to unity. Even bearing in
mind the relatively large uncertainties in the numerically obtained amplitudes,
this improvement is significant, at least for the $(4,3)$ and $(5,5)$ mode.
\begin{table}[t]
  \centering
  \caption{\label{tab:alpha}Deviations of the rescaling factor
          $\alpha$ introduced in Eq.~(\ref{eq:hybrid}) from unity
          for several multipoles and using the
          multipole expressions of Ref.~\cite{Kidder2007} or
          Ref.~\cite{Blanchet2008} which includes higher-order PN
          terms for subdominant modes.}
  \begin{tabular}{c|cccccc}
  \hline \hline
  $(l,m)$ & $(2,2)$ & $(3,3)$ & $(3,2)$ & $(4,4)$ & $(4,3)$ & $(5,5)$ \\
  \hline
  $\alpha-1$ (T1, \cite{Kidder2007})  & 0.047 &-0.071 &-0.065 &-0.108 & 0.251 & 0.712 \\
  $\alpha-1$ (T1, \cite{Blanchet2008})& 0.047 & 0.023 &-0.059 &-0.108 &-0.075 &-0.106 \\
  $\alpha-1$ (T4, \cite{Kidder2007})  & 0.048 &-0.071 &-0.065 &-0.107 & 0.251 & 0.712 \\
  $\alpha-1$ (T4, \cite{Blanchet2008})& 0.048 & 0.023 &-0.058 &-0.107 &-0.075 &-0.106 \\
  \hline \hline
  \end{tabular}
\end{table}

\section{Conclusions}
\label{sec:conclusions}

We have studied numerical simulations of a non-spinning black-hole binary
system with mass-ratio $q=4$ lasting about 11 orbits prior to coalescence.
The numerical uncertainties for phase and amplitude due to
discretization are $\Delta \phi \approx 0.6~{\rm rad}$ and
$\Delta A / A \approx 1~\%$ for the quadrupole mode through inspiral,
merger and ringdown when aligning the waveforms at the peak of the
amplitude of the (2,2) multipole. The phase error
is approximately constant, however, and we estimate the uncertainty
for the purpose of a PN comparison closer to $\Delta \phi \approx
0.11~{\rm rad}$. Numerical error due to wave extraction at finite
radii results in larger uncertainties for the amplitude of
about $\Delta A/A \lesssim 5~\%$ and a phase error
$\Delta \phi \leq 0.2~{\rm rad}$ up to merger and $0.5~{\rm rad}$ in the
late ringdown stage. Uncertainties for subdominant multipoles are larger;
approximately in proportion to the multipole index $m$ for the phase
and reaching $10~\%$ to $25~\%$ for higher-order modes in the amplitude.
We also observe agreement within numerical uncertainties
with an independent simulation of a $q=4$ binary obtained
with the {\sc BAM} code.

We have performed a matching to post-Newtonian predictions using the
Taylor T1 and T4 approximants and employing 3PN accurate expressions for
the GW multipoles. Our main results in this comparison can be summarised as
follows.

Using our numerical waveforms at finite extraction radii gives the misleading
impression that Taylor T4 produces better agreement than the T1 approximant.
Including subdominant multipoles with $l\ne m$ explicitly demonstrates the
internal inconsistency of the numerical results at finite (too small)
extraction radii: contrary to expectations the $(2,2)$ and $(3,2)$ as well
as the $(3,3)$ and $(4,3)$ multipoles are significantly out of phase.
This feature
improves when going to larger radii and disappears when results are extrapolated
to infinite radius. Subdominant modes thus provide valuable tests for the
internal consistency of numerical results, irrespective of whether they are
included in a comparison with post-Newtonian predictions or not.

By using extrapolated numerical results, we find the Taylor T1 approximant
to provide better agreement with numerical results:
using a matching window in the early inspiral,
the accumulated phase error at $M\omega_{22}=0.1$ is
$\Delta \phi \approx 0.35~{\rm rad}$ for T1 compared with $1.0~{\rm rad}$
for T4. For subdominant multipoles we obtain deviations of PN from NR
results which to good approximation
are proportional to the multipole index $m$.

The inclusion of higher-order PN terms in expressions for sub-dominant
multipoles in \cite{Blanchet2008} leads to improved amplitude agreement
with numerical results in our matching window covering approximately
the frequency range $0.05 \leq M\omega \leq 0.06$. Even bearing in mind the
relatively large numerical errors for the low amplitude modes, this
improvement is significant at least for the $(4,3)$ and the $(5,5)$ multipole.

\section*{Acknowledgments}
We thank Emanuele Berti for valuable discussions. C.F.S. and U.S.~acknowledge
support from the Ram{\'o}n y Cajal Programme of the Spanish Ministry
of Education and Science (MEC). U.S. acknowledges support
by FCT-Portugal through Project
No.~PTDC/FIS/098025/2008, by grants from the Sherman Fairchild Foundation
to Caltech and by NSF grants PHY-0601459, PHY-0652995, PHY-0960291.
D.M. acknowledges support by the
DFG Research Training Group 1523 ``Quantum and Gravitational Field''.
C.F.S. acknowledges support from a Marie Curie International Reintegration
Grant No.~MIRG-CT-2007-205005/PHY within the 7th
European Community Framework Programme (EU-FP7) and contracts
ESP2007-61712 (Spanish Ministry of Education and Science)
and No.~FIS2008-06078-C03-01/FIS (Spanish Ministry of Science and
Innovation).
This work was supported by allocations through the TeraGrid Advanced
Support Program under Grant Nos.~PHY-090003 and AST-100021 on the NICS
Kraken and SDSC Trestles clusters, by an allocation through
the Centro de Supercomputaci{\'o}n de Galicia (CESGA) under project
numbers ICTS-2009-120 and ICTS-CESGA-175
and Grant DECI-6 DyBHo by the EU-FP7 DEISA.
This work was supported in part by DFG grant SFB/Transregio~7 ``Gravitational
Wave Astronomy'' and the DLR (Deutsches Zentrum
f\"ur Luft und Raumfahrt).
{\sc BAM} simulations were performed at HLRB2 of LRZ Munich.



\section*{References}
\bibliographystyle{unsrt-notitle}

\end{document}